\documentclass[epj, nopacs]{svjour}
\usepackage{graphics}
\usepackage[numbers]{natbib}
\usepackage{amsmath}
\usepackage{hyperref}
\usepackage{breqn}
\usepackage{bm}
\usepackage{amssymb}
\usepackage{dcolumn}
\usepackage{gensymb}
\usepackage{multicol}

\usepackage[font=small,skip=1pt]{caption}

\usepackage[utf8]{inputenc}
\usepackage{newtxtext}
\usepackage[mathcal]{euscript}
\usepackage[normalem]{ulem}

\usepackage[export]{adjustbox}
\usepackage{float}
\usepackage{lipsum}

\usepackage{graphicx}
\usepackage[caption=false]{subfig}
\usepackage[usenames,dvipsnames]{xcolor}
\graphicspath{{figures/}}

\usepackage{hyperref}
\hypersetup{colorlinks=true,citecolor=blue,linkcolor=blue,urlcolor=blue}

\usepackage{diagbox}
\usepackage{booktabs}
\usepackage{tabularx}
\usepackage{amsmath,amssymb,amsfonts}
\usepackage{mathrsfs}
\usepackage{slashed}
\usepackage{bm}
\usepackage{multirow}

\usepackage[draft,commentmarkup=uwave]{changes} 
\definecolor{AllieViolet}{rgb}{0.5, 0.0, 1.0}
\definechangesauthor[color=green!50!black]{Wim}
\definechangesauthor[color=AllieViolet]{Allie}

\definechangesauthor[color=blue!50!black]{Alan}
\usepackage{tensor}
\usepackage{breqn}

\begin{document}
\title{Polarization options in inclusive DIS off tensor polarized deuteron}
\author{Wim Cosyn\inst{1}\thanks{email: wcosyn@fiu.edu} \and Brandon Roldan Tomei\inst{1,2}\thanks{email: brold003@fiu.edu} \and Alan Sosa\inst{1}\thanks{email: asosa90@fiu.edu} \and Allison Zec\inst{3}\thanks{email: allison.zec@unh.edu}} 
%
%
\institute{Department of Physics and Astronomy, Florida International University, Miami, FL 33199, USA \and Department of Physics, Florida State University, Tallahassee, FL 32306, USA \and Department of Physics and Astronomy, University of New Hampshire, Durham, NH 03824, USA}
%
\date{Received: date / Revised version: date}
%
\abstract{In the near future, the Jefferson Lab $b_1$ experiment will provide the second measurement of tensor polarized asymmetries in inclusive DIS on the deuteron.  In this asymmetry, 4 independent tensor polarized structure functions contribute.  This necessitates systematic approximations in the extraction of the leading twist structure function $b_1$ from a single tensor asymmetry measurement.  Contamination from higher twist structure functions and kinematic effects is discussed here. Using a deuteron convolution model, we quantify the systematic errors from these approximations for two different choices for the target polarization direction (momentum transfer, electron beam direction).  For Jefferson Lab 12 GeV kinematics, the systematic error turns out to be comparable between the two polarization options, while at higher $Q^2$ values the momentum transfer direction is preferred.
%
} 

\authorrunning{Cosyn, Roldan, Sosa, Zec}
 \titlerunning{Polarization options in inclusive DIS off tensor polarized deuteron}
\maketitle
\section{Introduction}

The $b_1$ experiment is an experiment to run in Hall C of Jefferson Lab (JLab) that aims to measure the leading-twist tensor polarized structure function (SF) $b_1$ of spin-1 hadrons, in this case the deuteron~\cite{Slifer:2013b1}. The $b_1$ SF can be extracted from an inclusive DIS measurement in the reaction
\begin{equation}
    e(k) + \vec{d}(p_d) \to e'(k') + X,
\end{equation}
where the deuteron is polarized and no electron polarization is required.  

The observable of interest  is the tensor polarized asymmetry
\begin{align}\label{eq:AT}
    &A^T = \frac{d\sigma(\Lambda_d=+1)+d\sigma(\Lambda_d=-1)-2d\sigma(\Lambda_d=0)}{d\sigma(\Lambda_d=+1)+d\sigma(\Lambda_d=0)+d\sigma(\Lambda_d=-1)}\,,\\
    &[-2 \leq A^{T}\leq 1]\nonumber
\end{align}
where the individual terms denote the differential cross section for the deuteron state with polarization $\Lambda_d$ along some fixed polarization direction (discussed in more later).

\begin{figure}[!htb]
    \centering
    \includegraphics[width=\linewidth]{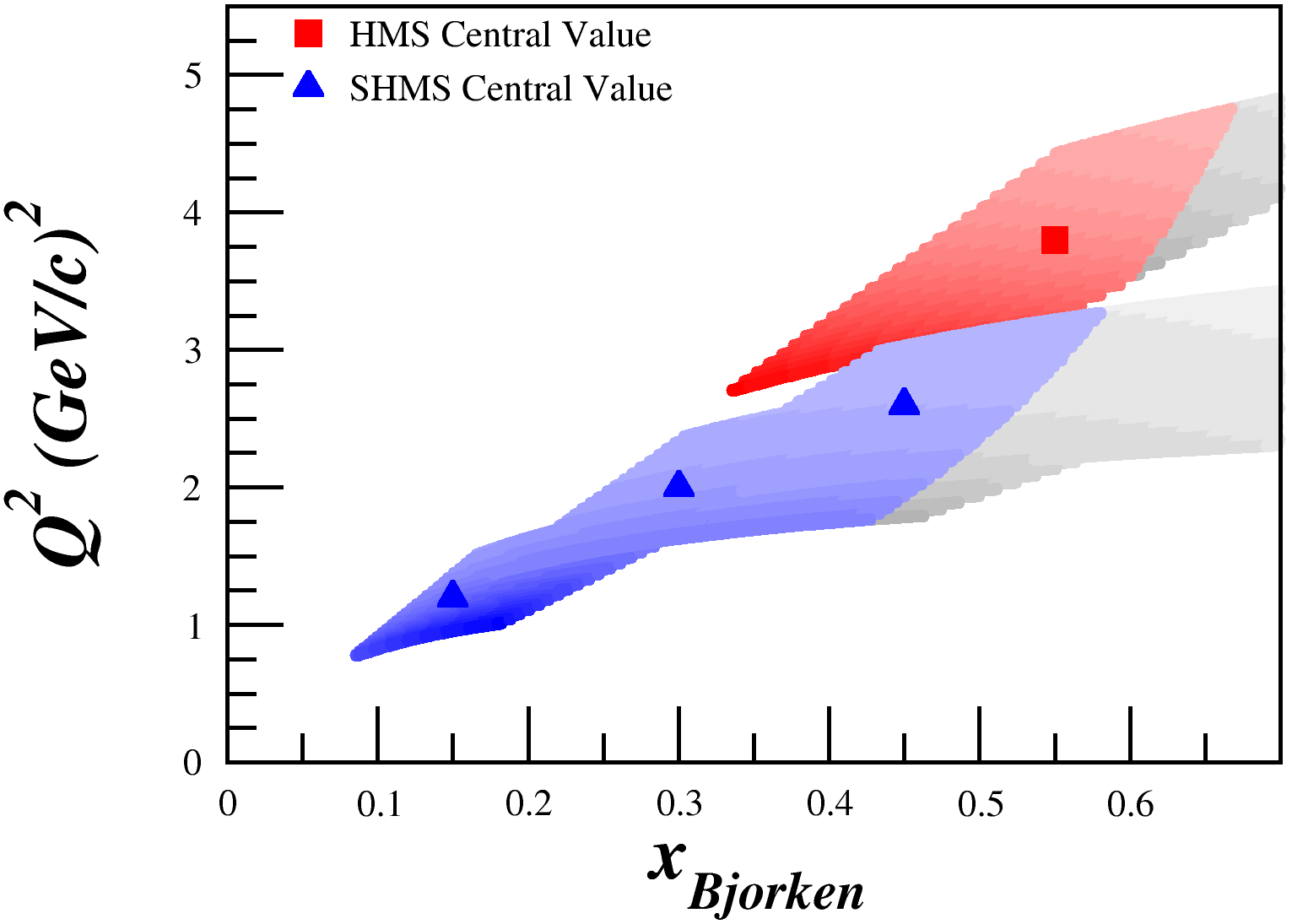}
    \caption{Kinematical coverage of the $b_1$ experiment over $x$ and $Q^2$. The blue region is the kinematical coverage of the left SHMS spectrometer in JLab Hall C, and the red area is that of the right HMS spectrometer. The grey regions are excluded by a cut of $W^2 > 1.85$ GeV$^2$. The darker areas of each region indicate kinematic points with higher rates \cite{Slifer:2013b1, b1-rates}.}
    \label{fig:b1-rates}
\end{figure}

\begin{figure}[hbt!]
    \centering
    \includegraphics[width=0.8\linewidth]{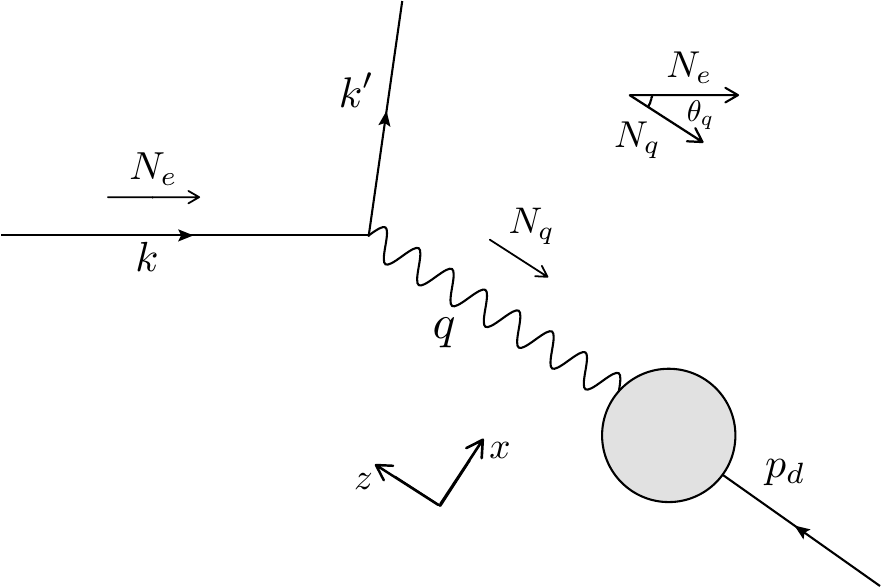}
    \caption{Unpolarized electron-polarized deuteron deep-inelastic scattering.  Incoming (outgoing) electron have momentum $k$ ($k'$), the deuteron has momentum $p_d$. $N_e$ ($N_q$) denote the deuteron polarization direction considered here, along the electron beam (virtual photon direction).}
    \label{fig:DIS}
\end{figure}

As shown in Fig.~\ref{fig:b1-rates}, the JLab experiment will cover DIS kinematics $0.8~\text{GeV}^2~< Q^2 < 5.0~\text{GeV}^2$ and $0.16 < x < 0.49$, with $0<x<2$ the rescaled Bjorken scaling variable and $Q^2$ the momentum transfer squared.  Figure \ref{fig:DIS} depicts the reaction and specifies the momenta of the initial and final-state particles, were the exchanged virtual photon carries momentum transfer $q=k-k'$.

The collinear partonic distribution function (usually denoted $\delta_T q$ or $f_{1LL}=-\tfrac{2}{3}\delta_T q$~\cite{Kumano:2024fpr}) that appears in the $b_1$ SF has a probabilistic interpretation in the parton model~\cite{Hoodbhoy:1988am}.  It is linked to the difference of quark (longitudinal momentum) distributions in the deuteron with polarization $\Lambda_d =+1$ and $\Lambda_d=0$:
\begin{equation}\label{eq:partb1}
    b_1 \propto [q^{+1}(x)-q^{0}(x)]\,.
\end{equation}
For the proton-neutron ($pn$) component of the deuteron, a non-zero $b_1$ SF is linked to the non-central (tensor) part of the nucleon-nucleon interaction.  As such $b_1$ provides a measure of how nuclear interactions affect collinear partonic structure.  In a convolution model~\cite{Frankfurt:1983qs,Hoodbhoy:1988am,Khan:1991qk,Cosyn:2017fbo}, $b_1$ is non-zero only in the presence of a deuteron D-wave ($L=2$).  $f_{1LL}$, the partonic distribution function that contributes to $b_1$, enters in the quark total and orbital angular momentum sum rules for spin-1 hadrons~\cite{Cosyn:2019aio}.

In the past, HERMES has provided the only measurement of $A^T$\footnote{There and elsewhere denoted as $A_{zz}$.} in the DIS region and provided an extraction of $b_1$~\cite{HERMES:2005pon}.  The tensor asymmetries and associated $b_1$ values are typically small for inclusive DIS measurements.  This is due to i) the $D$-wave being a small component of the deuteron wave function at internal $pn$-momenta $\leq \text{few}~ 100~\text{MeV}$,  ii) the inclusive measurement averaging over all initial deuteron configurations.   Conventional nuclear convolution models are not able to explain the HERMES result, with calculations typically smaller than the measured values~\cite{Khan:1991qk,Cosyn:2017fbo}.  This could be due to physics not included in the convolution models (non-nucleonic components in the deuteron for instance~\cite{Miller:2013hla}), but could also be due to the systematic errors associated with approximations in the extraction procedure. The latter effect is discussed and quantified in more detail in this article.

As mentioned earlier, Jefferson Lab has an approved experiment to measure the inclusive tensor asymmetry, both in the quasi-elastic ($x\geq1$) and deep inelastic region. The experiment aims to extract $b_1$ over a similar kinematic region as the HERMES data. While the HERMES reported a 44\% relative uncertainty for their $A^T$ measurement, the proposal for the $b_1$ experiment at Jefferson Lab calls for a measurement capable of achieving 9.2\% relative uncertainty on $A^T$ \cite{HERMES:2005pon, b1}. In this light, we discuss in this article some of the approximations that are needed to extract $b_1$ from the tensor asymmetry measurement, especially with regard to the freedom that one has in picking the direction in which to polarize the deuteron target.  As it is not trivial to change the polarization direction once the target is operated, carrying out this study is of relevance and importance to the planned measurement.  

In Sec.~\ref{sec:pol} we discuss how the polarization for spin-1 targets is quantified through variables that enter theoretical calculations and experimental polarimetry of spin-1 ensembles.  Next, in Sec.~\ref{sec:xsection_asymm}, we discuss the general form of the inclusive DIS cross section and tensor asymmetry for spin-1 targets, while in Sec.~\ref{sec:directions} we cover how the polarization direction affects the asymmetry and what approximations are needed to obtain $b_1$ from a single tensor asymmetry.  In Sec.~\ref{sec:errors} we estimate the systematic error that follows from the used approximations in the $b_1$ extraction at JLab kinematics.  Conclusions are stated in Sec.~\ref{sec:concl}.

\section{Deuteron Polarization}
\label{sec:pol}
The $b_1$ experiment uses a solid polarized deuteron target, which will be achieved using a cryogenic target material of deuterated ammonia (ND$_3$) which is doped with free radicals as well as the process of dynamic nuclear polarization (DNP). The usage of DNP introduces a number of technical requirements for the target such as the usage of cryogenics to hold the target at a low temperature (typically about 1 K), and a magnetic holding field (typically either 2.5 T or 5 T) \cite{CrabbMeyer}. Due to the Zeeman splitting of nuclear spin states, at low temperature nuclear spin states will fill up preferentially at low temperature according to Boltzmann statistics. This effect can be quantified by the nuclear spin polarization $n_I = \langle I_z \rangle / I$ where $I$ is the total population, and $I_z$ is the population of nuclei with spin aligned along B-field axis $Z$. 

For a spin-1 particle such as a deuteron, the Zeeman splitting creates three energy sublevels each corresponding to the $\Lambda_d=+1, 0$, and -1 spin states. The ``vector'' polarization $\mathcal{P}$ is then:
\begin{equation}\label{eq:vectorpol}
    \mathcal{P} = n_{+} - n_{-}
\end{equation}
where $n_{+}$, $n_{-}$ and $n_{0}$ ($n_+ + n_0 + n_-=1$) are the relative populations of the spin $\Lambda_d=1$ state, the spin $\Lambda_d=-1$ state, and the spin $\Lambda_d=0$ state respectively. For deuterons in a magnetic field $B$ and at temperature $T$ the thermal equilibrium vector polarization is

\begin{equation}
    \mathcal{P}^{(\text{TE})} = \frac{4 \tanh\left( \frac{g_d \mu_N B}{2 k_B T} \right)}{3 + \tanh^2\left( \frac{g_d \mu_N B}{2 k_B T} \right)},
\end{equation}
where $B$ is the magnetic holding field, $T$ is the target temperature, $g_d$ is the deuteron g-factor ($g_d\simeq 0.857$) and $\mu_N$ is the nuclear magneton.

At a magnetic field of $B=5$~T and at a temperature of $T=1$~K the deuteron polarization is still quite small, only about 0.1\%. However, at the same field and temperature the electron polarization is quite high, over 99\% polarized. The nuclear polarization can then be enhanced by driving electron spin transitions in the target, and which causes the unpaired electrons from the free radicals to transfer their spins to the nuclei. The spin transitions are driven with the application of microwaves tuned to the electron spin transition frequency (approximately 140 GHz at 5 T) during this process of microwave ``enhancement'' of the polarization, the nuclear spin polarization builds up according to a number of factors including the density of free radicals in the target sample, the microwave power, and the magnetic field uniformity within the target sample.

In addition to vector polarization, enhancement of deuterons also drives an increase in ``tensor'' polarization $\mathcal{Q}$ \cite{MEYER1986574} which is
\begin{equation}\label{eq:tensorpol}
    \mathcal{Q} = n_{+} + n_{-} - 2n_{0}.
\end{equation}
As discussed in Sec.~\ref{sec:xsection_asymm}, the inclusive cross section with an unpolarized electron beam is only sensitive to tensor polarization $\mathcal{Q}$, not vector polarization $\mathcal{P}$.

From the theory side, the deuteron ensemble with different polarization states is described by a spin-1 density matrix.  Here, we use the covariant density matrix, which can be parameterized in terms of an axial 4-vector $s^\mu$ and a traceless and symmetric 4-tensor $t^{\mu\nu}$
\begin{equation}
    \rho^{\mu\nu} = \frac{1}{3}\left( -g^{\mu\nu}+\frac{p_d^\mu p_d^\nu}{M_d^2}\right) + \frac{i}{2M_d}\epsilon^{\mu\nu\rho\sigma}p_{d,\rho}s_\sigma -t^{\mu\nu},
\end{equation}
where $M_d$ is the deuteron mass, $p_d$ the deuteron momentum and $s^\mu (t^{\mu\nu})$ are the polarization vector (tensor) and $\epsilon^{0123}=1$.
For more details on $\rho^{\mu\nu}$, see Ref.~\cite{Cosyn:2020kwu}.

Pure deuteron states are characterized by a polarization direction $N^\mu$ [$(Np_d)=0$] and a polarization state $\Lambda_d=\pm 1,0$, with total angular momentum states quantized along $N$.  In the deuteron rest frame $N^\mu$ reduces to the spatial direction used to polarize the deuteron ensemble $N^\mu = (0,\bm N)$.  For pure states, the density matrix parameters take the following form.  The axial 4-vector can be expressed as 
\begin{equation}
    s^\mu(N,\Lambda_d)=\Lambda_d N^\mu, \quad  (p_d N)=0, \quad N^2=-1.
\end{equation}
For pure states, the symmetric and traceless 4-tensor can be expressed as 
\begin{align}\label{eq:spin1_tensor_pol}
    t^{\alpha\beta}(N,\Lambda_d)= \frac{1}{6}(g^{\alpha\beta}&-\frac{{p_d}^\alpha {p_d}^\beta}{{p_d}^2}+3N^\alpha N^\beta)\nonumber\\ &\times 
    \begin{cases}
        +1 &\Lambda_d = \pm 1\\
        -2 &\Lambda_d = 0
    \end{cases}.
\end{align}
In the deuteron rest frame the eigenvalues of $ \rho^{\alpha\beta}$ (being $n_+,n_0,n_-$) can be expressed entirely using the degrees of polarization $\mathcal{P}$ and $\mathcal{Q}$, see Eqs.~(\ref{eq:vectorpol}) and (\ref{eq:tensorpol}).

\section{Cross section and tensor polarized asymmetry}
\label{sec:xsection_asymm}

In the inclusive DIS cross section, the tensor degrees of freedom present in the spin-1 density matrix lead to four additional tensor polarized structure functions in comparison with the spin-1/2 case.
For unpolarized electron scattering\footnote{In the case of polarized electron beams, two additional structure functions appear ($g_{1}$ and $g_{2}$), which couple to deuteron vector polarization only.}, this results in the following decomposition of the cross section~\cite{Cosyn:2020kwu}

\begin{align}
\label{eq:cross}
    \frac{d\sigma}{dxdQ^2} =\frac{\pi y^2 \alpha^2}{Q^4(1-\epsilon)} &\left[ F_{[UU,T]}+\epsilon F_{[UU,L]} \right.\nonumber\\
    &\left.+T_{LL}(F_{[UT_{LL},T]}+ \epsilon F_{[UT_{LL},L]})\right.
    \nonumber\\
    &\left.+\sqrt{2\epsilon(1+\epsilon)} \, T_{LT}\cos\phi_{T_L} F_{[UT_{LT}]}\right.\nonumber\\
    &\left.
    +\epsilon \,T_{TT}\cos2\phi_{T_T} \, F_{[UT_{TT}]} \right]
\end{align}
Here, $\alpha$ is the electromagnetic fine structure constant.  The kinematical invariants Bjorken $x$, inelasticity $y$ and Rosenbluth variable $\epsilon$ are defined as follows
\begin{subequations}\label{eq:kinematics}
    \begin{align}
        x_d=\frac{-q^2}{2(p_dq)}, \qquad (0<x_d<1), \\
        y=\frac{(p_d q)}{(p_d k)}, \qquad (0<y<1),\\
        \epsilon = \frac{1-\gamma^2y^2/4}{1-y+y^2/2+\gamma^2y^2/4},
    \end{align}
\end{subequations}
In Fig.~\ref{fig:b1-rates} and what follows we use the rescaled
\begin{equation}
    x = 2x_d, \qquad (0<x<2),
\end{equation}
which permits more natural comparison with the Bjorken-$x$ variable on the nucleon (i.e. $x\approx 1$ corresponds to quasi-elastic scattering).  
The variable $\gamma$ is defined as
\begin{equation}\label{eq:gamma}
    \gamma=\frac{M_dQ}{(p_dq)}= \frac{2x_dM_d}{Q}.
\end{equation}

In the deuteron rest frame (RF) depicted in Fig.~\ref{fig:DIS}, the polarization parameters appearing in Eq.~(\ref{eq:cross}) can be identified with components of the polarization tensor of Eq.~(\ref{eq:spin1_tensor_pol}):
\begin{subequations}\label{eq:tensor_RF}
\begin{align}
    T_{LL} = t^{zz}[\text{RF}],\\
    T_{LT}\cos \phi_{T_L} = t^{xz}[\text{RF}],\\
    T_{TT}\cos 2\phi_{T_T} = t^{xx}[\text{RF}]-t^{yy}[\text{RF}].
\end{align}    
\end{subequations}
For details on the construction and interpretation of these parameters as Lorentz invariants, see Refs.~\cite{Cosyn:2019hem,Cosyn:2020kwu}.

We follow the notation used in \cite{Cosyn:2020kwu} to identify the structure functions (SF) indexed  with the  different electron, deuteron, and virtual photon polarizations:
\begin{equation}\label{eq:strucPol}
 F_{[\text{electron deuteron,photon}]}(x,Q^2).
\end{equation}
The dependence on $x$ and $Q^2$ is usually suppressed and implicit in any of the other equations.
The structure functions contain information the high-energy DIS process and the internal deuteron structure probed in the reaction.

For the unpolarized structure functions, we have the following relations with $F_1, F_2$
\begin{align}\label{eq:ogStruct}
     F_{[UU,L]} &= (1+\gamma^2)\frac{F_2}{x_d}-2 F_1\,,\nonumber\\
F_{[UU,T]}  &= 2F_1.
\end{align}

The tensor polarized structure functions were originally introduced in~\cite{Hoodbhoy:1988am} as $b_1$ to $b_4$ using a different decomposition of the hadronic tensor.  The linear relations with those appearing in Eq.~(\ref{eq:cross}) is given by
\begin{subequations}\label{eq:F_b_relation}
\begin{align}
 &F_{[U T_{LL},T]}   =
-\left[2(1+\gamma^2)b_1-\frac{\gamma^2}{x_d}\left(\frac{1}
{ 6
}
b_2-\frac{1}{2}b_3\right)\right ] \,,\\
    &F_{[U T_{LL},L]}   =
\frac{1}{x_d}\left[2(1+\gamma^2)x_d b_1\right.\\
& \qquad\qquad-(1+\gamma^2)^2
\left(\frac { 1 } {
3}b_2+b_3+b_4\right)\nonumber\\
& \qquad\qquad\left.-(1+\gamma^2)\left(\frac{1}{3}b_2-
b_4\right)-\left(\frac{1}{3}b_2-b_3 \right)\right ]
\,,\\
  &F_{[U T_{LT}]}
  =
-\frac{\gamma}{2x_d}\left[(1+\gamma^2)\left(\frac{1}{3}
b_2-b_4\right)\right.\nonumber\\
&\qquad\qquad\qquad\left. +\left(\frac{2}{3}b_2-2b_3\right)\right]\,,\\
 &F_{[U T_{TT}]}
  = - \frac{\gamma^2}{x_d}
\left(\frac{1}{6}b_2-\frac{1}{2}b_3 \right)\,.\label{eq:inclusivestruc}
    \end{align}    
\end{subequations}

The SF used here have the advantage that the geometric dependence on the different tensor polarization parameters is more clear, while the $b_1-b_4$ have a more clear interpretation in the parton model.  
In the Bjorken limit ($Q^2 \to \infty$, $x$ fixed), $b_1$ and $b_2$ obey a Callan-Gross-like relation
\begin{equation}\label{eq:b1_CG}
b_2 = 2x_d b_1 = x b_1\,. 
\end{equation}
$b_1$ and $b_2$ are leading twist, while $b_3$ and $b_4$ are higher twist.  For the SF used here, Eqs.~(\ref{eq:F_b_relation}) imply that $F_{[U T_{LL},L]}$ and $F_{[U T_{LL},T]}$ are leading twist, while $F_{[U T_{LT}]}$ and $F_{[U T_{TT}]}$ are higher twist.

By using Eqs.~(\ref{eq:cross}) (see Ref.~\cite{Cosyn:2019hem}), we can write the asymmetry as
\begin{align}\label{eq:AT_SF}
        A^T = \frac{1}{\mathcal{Q}} \frac{2}{F_{UU,T}+\epsilon \, F_{UU,L} }\, \left[ T_{LL}(F_{U T_{LL},T}+\epsilon \, F_{U T_{LL},L}) \right. \nonumber \\
        \left. + T_{LT}\cos\phi_{T_L} \sqrt{2\epsilon(1+\epsilon)}\, F_{[U T_{LT}]} + T_{TT}\cos2\phi_{T_T}\, \epsilon \, F_{[U T_{TT}]}\right]
\end{align}
where it is understood that $\mathcal{Q}$ is the combination of Eq.~(\ref{eq:tensorpol}) made using the eigenvalues of the density matrix of the deuteron ensemble that is being considered.

 Experiments wishing to extract $b_1$ have to do so from a tensor asymmetry measurement.  One tensor asymmetry measurement is, however, not sufficient to uniquely extract one SF as four unknown structure functions enter in the numerator of the asymmetry.  In order to extract the $b_1$ structure function either multiple measurements with different polarization directions $\bm N$ (which results in different values for the tensor polarization parameters)  and/or $\epsilon$ are needed or an approximation has to be used to eliminate some of the unknowns.   As performing measurements with different polarization directions is technically challenging and results in larger beam time needs, the HERMES measurement and the upcoming JLab experiment use a single tensor asymmetry measurement with approximations to extract $b_1$.  These approximations result in additional systematic errors for the extraction.  However, for a single measurement there is still the question which polarization direction $\bm N$ could be the most optimal and yields the best approximation.  

 \section{Polarization direction considerations}
\label{sec:directions}

In this section, we discuss possible approximations that have or can be used to extract $b_1$ from a single $A^{T}$ measurement. While from a theory viewpoint, we could analyse Eq.~(\ref{eq:AT_SF}) for any direction $\bm N$, the experimental setup has real constraints in which directions the deuteron can be polarized.

\begin{figure}[!hbt]
    \centering
    \includegraphics[width=\linewidth]{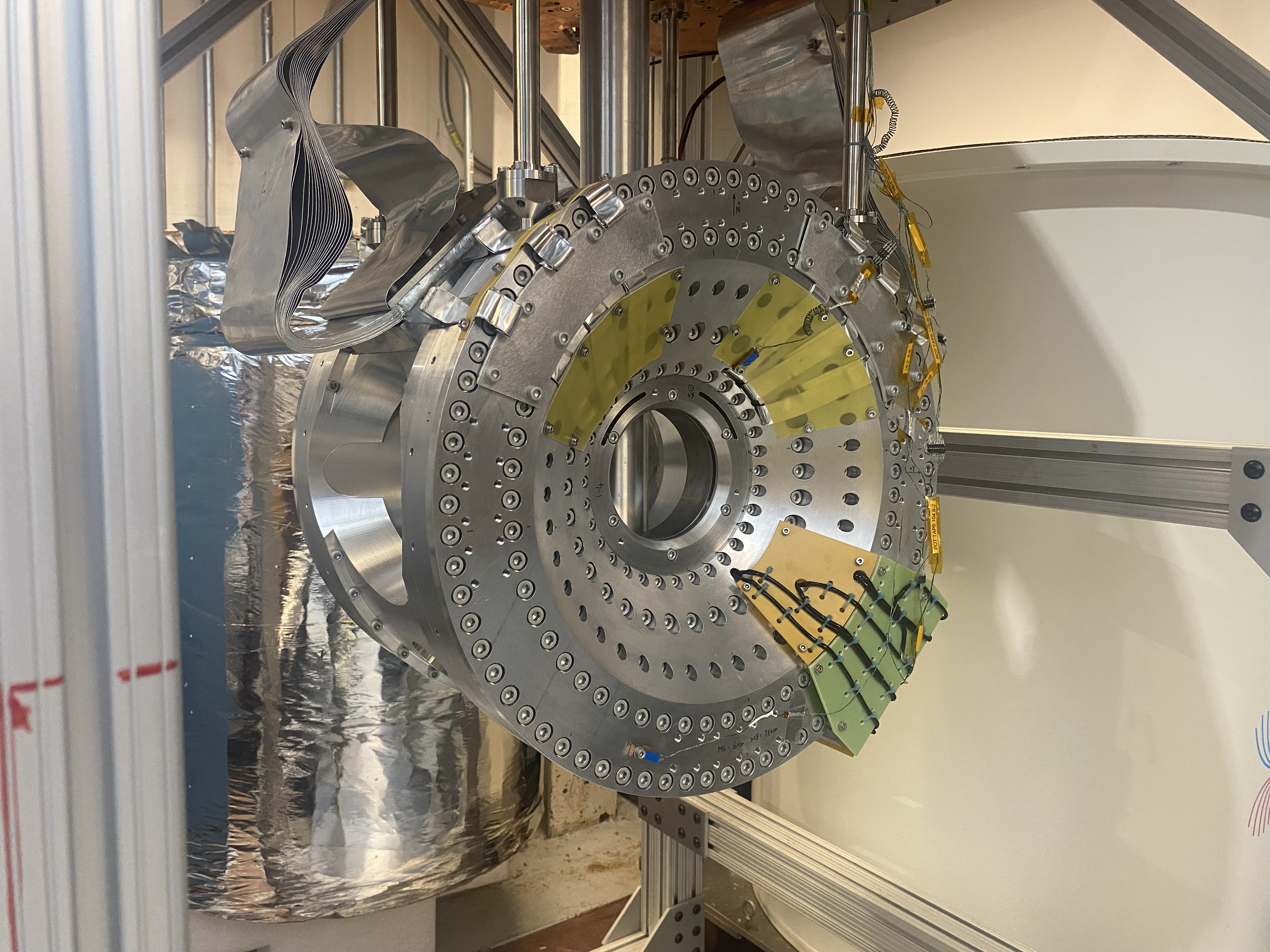}
    \caption{A photograph of the 5 T superconducting magnet to be used for the polarized target field for the $b_1$ experiment.}
    \label{fig:hall-c-magnet}
\end{figure}

Because of the need for a cryogenic target in a high magnetic field with microwave transmission for polarization enhancement, the maintenance of the target can get quite complex for any practical measurement of $A^T$. This complexity has the effect of practically limiting the target and detector configurations that an inclusive measurement can be designed to have. In particular, Jefferson Lab's most recent polarized target run employed a 5 T superconducting magnet, which has an opening bore of $\pm$35\degree{} and is designed to produce a field direction either longitudinal or transverse to the beamline (as seen in Fig. \ref{fig:hall-c-magnet}). In the longitudinal configuration, which is planned to be used for the $b_1$ experiment, the deuteron is polarized along the beam momentum~\cite{magnet-convo}. Rotating the magnet anywhere within its $\pm$35\degree{} opening bore angle requires the use of a magnetic chicane to ``pre-steer'' the electron beam in order to correctly reproduce the longitudinal-field kinematics, which can complicate the running of the experiment. However, running with a longitudinal field also creates a nonzero angle $\theta_q \neq 0$ between the target polarization direction and the momentum transfer direction $\bm N_q$, which as shown below results in a more complicated relation between the asymmetry $A^T$ and $b_1$. 

The tensor asymmetry of Eq.~(\ref{eq:AT_SF}) depends on the polarization direction of the deuteron through the tensor polarization parameters of Eqs.~(\ref{eq:tensor_RF}).  
Two natural polarization directions we can consider is i) polarizing along the virtual photon direction $\bm N\equiv \bm N_q$ and ii) polarizing along electron beam direction $\bm N\equiv\bm N_e$. For a deuteron polarized along the virtual photon direction, using Eqs.~(\ref{eq:spin1_tensor_pol}) and (\ref{eq:tensor_RF}), the tensor polarization parameters for $\Lambda_d=+1$ reduce to
\begin{subequations}\label{eq:polparams_photon}
\begin{align}
    &T_{LL} = \frac{\mathcal{Q}}{3},\\
    &T_{LT}\cos\phi_{T_L} = T_{TT}\cos2\phi_{T_T}= 0.\\
    &(\bm N=\bm N_q) \nonumber
\end{align}    
\end{subequations}
These values result in only two tensor polarized structure functions appearing in the asymmetry of Eq.~(\ref{eq:AT_SF}), reducing the complexity of the asymmetry.  For this choice of polarization direction we have that $\mathcal{P}$ and $\mathcal{Q}$ introduced in Sec.~\ref{sec:pol} are related to the tensor polarization parameters as
\begin{subequations}
\begin{align}
    &\mathcal{P} = S_L = s^z[\text{RF}],\\
    &\mathcal{Q} = 3\,T_{LL}.
\end{align}    
\end{subequations}

\begin{table*}[ht]
    \centering
    \renewcommand{\arraystretch}{1.3}
    \begin{tabular}{ |l|l|l|l|l|c|}
 \hline
  Twist &$\bm N_q$ &$\bm N_e$ & $\gamma $&  SF & Approximation used for $b_1/(F_1A^T)$\\
 \hline
 Leading& \checkmark & \checkmark   & 0 &  $b_2=xb_1$ &    $ -\frac{3}{2} $\\ 
 \hline
 \multirow{2}{*}{Kinematical Higher} &  \multirow{2}{*}{\checkmark} & & \multirow{2}{*}{Finite}   & $b_2=xb_1$,&    \multirow{2}{*}{$\frac{9(1+\epsilon \gamma^2)}{-6-5\gamma^2
    -2\epsilon\gamma^4}$} \\
    &&&&$b_3=b_4=0$&\\
 \hline
  \multirow{2}{*}{Kinematical Higher} & &  \multirow{2}{*}{\checkmark}  &  \multirow{2}{*}{Finite}    & $b_2=xb_1$,&   \multirow{2}{*}{$\frac{18(1+\epsilon \gamma^2)}{[(3\cos^2\theta_q-1)]([-6-5\gamma^2]
    -2\epsilon\gamma^4) - [\cos\theta_q\sin\theta_q]\sqrt{2\epsilon(1+\epsilon)}(9\gamma+3\gamma^3) - [\sin^2\theta_q] \epsilon (3\gamma^2)} $}  \\
&&&&$b_3=b_4=0$&\\
 \hline
\end{tabular}
    \caption{Summary of the different approximations used to relate $b_1$ to $A^T$.  The first row is the Bjorken limit approximation that for both polarization along the electron or photon direction leads to Eq.~(\ref{eq:AT_approx}). The second and third row keep contributions proportional to $\gamma$ but put the higher twist $b_3=b_4=0$.  All approximations use the tensor Callan-Gross relation.}
    \label{table:Approxb1}
\end{table*}

For a deuteron polarized along the electron beam direction, the polarization parameters take the form
\begin{subequations}\label{eq:polparams_angle}
\begin{align}
    &T_{LL} = \frac{1}{6}(3\cos^2\theta_q-1)\mathcal{Q} = \frac{3\cos 2\theta_q+1}{12}\,\mathcal{Q},\\
    &T_{LT}\cos\phi_{T_L}  = -\frac{1}{2}\cos\theta_q\sin\theta_q = -\frac{\sin 2\theta_q}{4}\,\mathcal{Q},\\
    &T_{TT}\cos2\phi_{T_T}  = \frac{\sin^2\theta_q}{2} = \frac{1-\cos2\theta_q}{4}\,\mathcal{Q},\\
    &(\bm N=\bm N_e) \nonumber
\end{align}    
\end{subequations}

where $0\leq \theta_q \leq \pi$ is the angle between virtual photon and electron beam in the target rest frame
\begin{equation}
    \cos\theta_q = \bm N_q \cdot \bm N_e,
\end{equation}
see Fig.~\ref{fig:DIS}.  This rest frame angle can also be expressed using invariants:
\begin{subequations}\label{eq:angle}
\begin{align}
    \cos \theta_q &= \frac{1 + \gamma^2 y/2}{\sqrt{1 + \gamma^2}},\\
    \sin \theta_q &= \frac{\gamma \sqrt{1 - y - \gamma^2 y^2/4}}{\sqrt{1 + \gamma^2}}.
\end{align}    
\end{subequations}

In the Bjorken limit, we have $\gamma \to 0$, and Eqs.~(\ref{eq:angle}) reduce to  $\cos\theta_q=1, \sin\theta_q=0$.  This means that in the Bjorken limit the polarization parameters for polarization along the virtual photon direction (Eqs.~(\ref{eq:polparams_photon})) or the electron beam direction (Eqs.~(\ref{eq:polparams_angle})) become identical.    Using the scaling limit Callan-Gross relations of Eq.~(\ref{eq:b1_CG}) and $F_2=2x_d F_1 = xF_1$,
 Eqs.~(\ref{eq:F_b_relation}) reduce in the Bjorken limit to
\begin{align}\label{eq:approxStruct}
& F_{[U T_{LL},T]} = -2b_1,\\
    &F_{[U T_{LL},L]}  = F_{[U T_{LT}]}  = F_{[U T_{TT}]}  =   F_{[UU,L]}  = 0.    
\end{align}
Consequently, for both polarization considerd here ($\bm N=\bm N_e,\bm N_q$) the tensor asymmetry simplifies in the Bjorken limit to 
\begin{equation} \label{eq:AT_approx}
    A^{T} = -\frac{2}{3}\frac{b_1}{F_1}.
\end{equation}
This is the relation that was used in the $b_1$ extraction of the HERMES result~\cite{HERMES:2005pon}.

The HERMES and JLab measurements have moderate $Q^2 \lesssim 5~\text{GeV}^2$ and are thus not close the Bjorken limit.  For reference, for the two $Q^2$ values we show in the plots in this paper, we have at $x=0.5$ that $\gamma=0.66~(Q^2=2~\text{GeV}^2)$ and $\gamma= 0.30~(Q^2=10~\text{GeV}^2)$. For this reason, we can attempt to relax some of the limits taken to arrive at Eq.~(\ref{eq:AT_approx}). Table \ref{table:Approxb1} shows a summary of the different limits considered in this analysis and the relation between $A^T$ and $b_1$ they result in.  Next to the Bjorken limit case of Eq.~(\ref{eq:AT_approx}) (first row), we retain the finite value for $\gamma$ in the other two approximations that we use (second and third row). Here, we still impose $b_3,b_4=0$ for the higher twist SF and use the tensor Callan-Gross relation, as otherwise we still have too many unknowns to solve for $b_1$ from a single asymmetry.  This approach can be considered including kinematical higher twist corrections, while still neglecting (out of necessity) the dynamical higher twist contributions.

\section{Comparing directions: systematic error estimation}
\label{sec:errors}

 For kinematics away from the Bjorken limit, using the approximate  relations in the right column of Table~\ref{table:Approxb1} to extract $b_1$ from data (measured tensor asymmetries $A^T$ and $F_1$ from unpolarized measurements) results in an extracted $b_1$ value that is different from the \emph{true} $b_1$.  The approximations lead to the introduction of an additional systematic error.  This systematic error can be checked and quantified using a model calculation, where both the true and extracted $b_1$ values can be computed and compared within the same framework.  Within the model, we can compute the structure functions appearing in Eq.~(\ref{eq:AT}) (and thus also the $b_{1-4}$) and the tensor asymmetry using Eq.~(\ref{eq:AT_SF}).  Next, we can extract $b_1$ using the approximations listed in Table~\ref{table:Approxb1} and compare with the original model calculation of $b_1$, referred to as the \emph{true} $b_1$. Through Eq.~(\ref{eq:AT}), we can disentangle the different contributions to the asymmetry: the different terms, but also within each term the effect of polarization, electron kinematics and additional structure functions.  We can study the dependence on model inputs (nucleon structure functions, deuteron wave function) and kinematics ($Q^2,x$ dependence) to investigate the robustness of the results.

We use a standard convolution model~\cite{Frankfurt:1983qs,Khan:1991qk,Cosyn:2017fbo} that calculates the proton--neutron component contribution to the deuteron structure functions. In the convolution model, the unpolarized and tensor polarized deuteron SF are computed through a convolution of unpolarized inclusive nucleon SF with deuteron light-front densities, see Ref.~\cite{Cosyn:2017fbo} for details of the formalism.  For the nucleon SF that enter the calculations, we compare both direct data-driven parametrizations of $F_{2N}$~\cite{Bodek:1979rx} (which include resonance contributions) and parametrizations based on the leading order relation with the unpolarized partonic distribution function (pdf), using the MSTW2008~\cite{Martin:2009iq} and CTEQ~\cite{Lai:1994bb} pdfs.  The deuteron light-front densities are matched to non-relativistic wave functions using the procedure detailed in~\cite{Cosyn:2020kwu} and we compare deuteron wave functions from Argonne V18~\cite{Wiringa:1994wb}, CDBonn~\cite{Machleidt:2000ge} and Paris~\cite{Lacombe:1980dr} parametrizations.  
We want to stress that no non-nucleonic components are included in the current analysis, for a study of these see~\cite{Miller:2013hla}.

\begin{figure}[hbt!]
    \centering
    \includegraphics[width=0.95\linewidth,center]{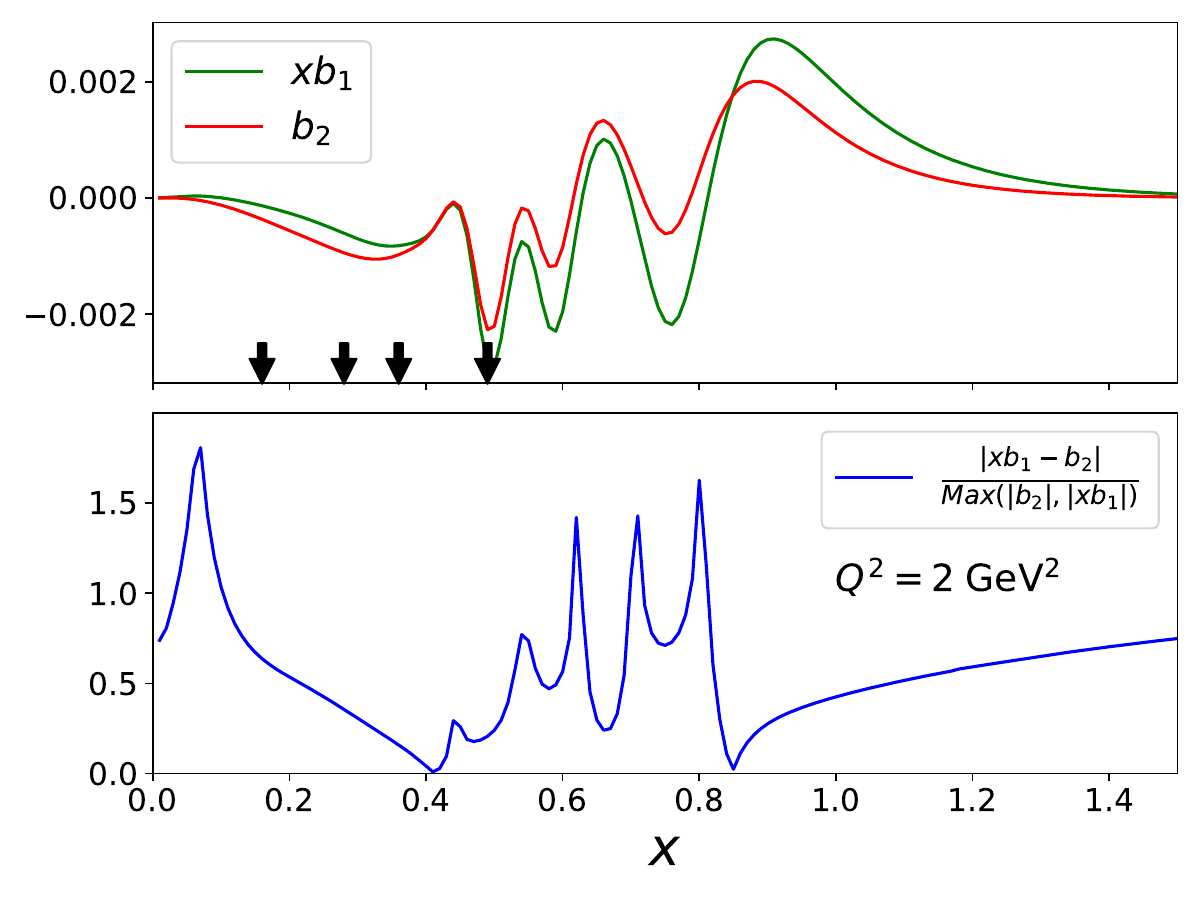}
    \includegraphics[width=0.95\linewidth,center]{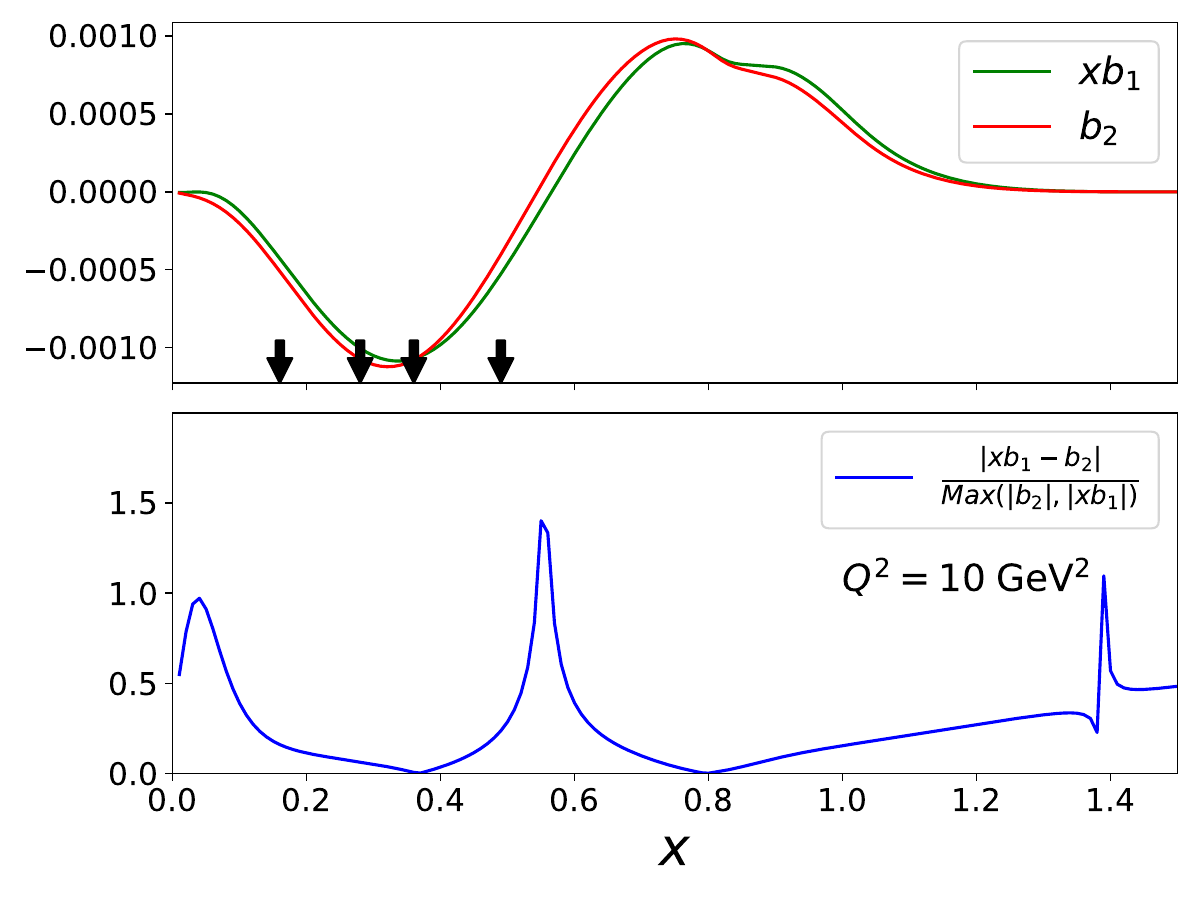}
    \caption{Validity of Callan-Gross relation at $Q^2=2~\text{GeV}^2$ (2 top panels) and $10~\text{GeV}^2$ (bottom 2 panels).  Top panels for each $Q^2$ show the absolute value of the relative difference between $b_2$ and $xb_1$.  This would be identically zero if the Callan-Gross relation is exactly valid.  The convolution model calculation uses the SLAC nucleon structure functions and Paris deuteron wave function. Black arrow indicate the central $x$-values for the JLab experiment, see Fig.~\ref{fig:b1-rates}.}
    \label{fig:cgQ2}
\end{figure}

\begin{figure*}[hbt!]
\centering
\begin{multicols}{2}
    {\includegraphics[width=0.95\linewidth,center]{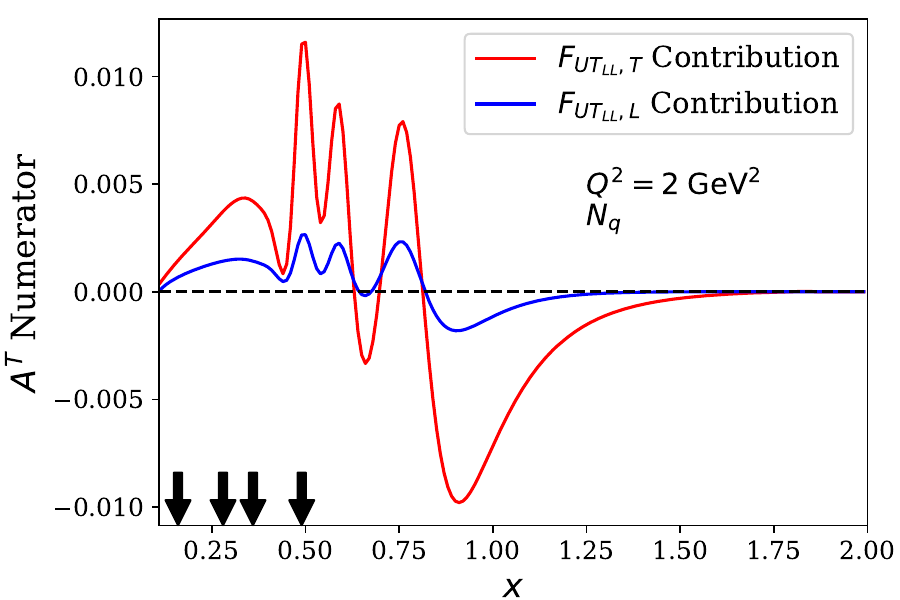}\par}
    {\includegraphics[width=0.95\linewidth,center]{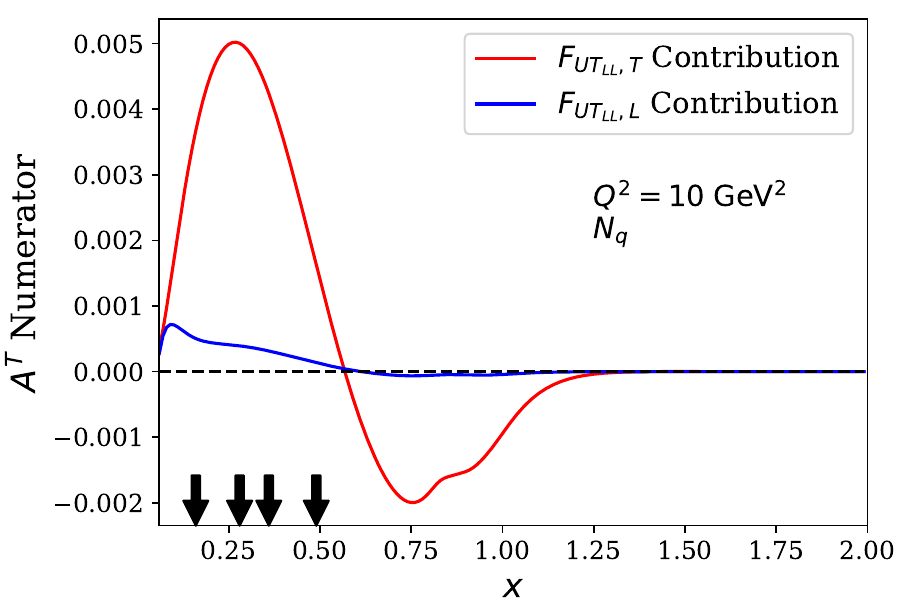}\par}
\end{multicols}
\begin{multicols}{2}
    {\includegraphics[width=0.95\linewidth,center]{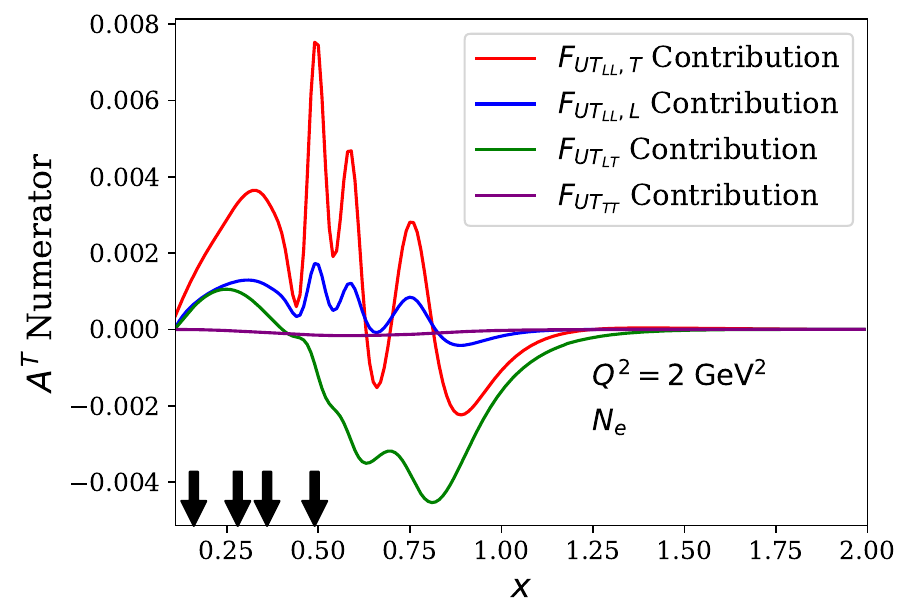}}\par
    {\includegraphics[width=0.95\linewidth,center]{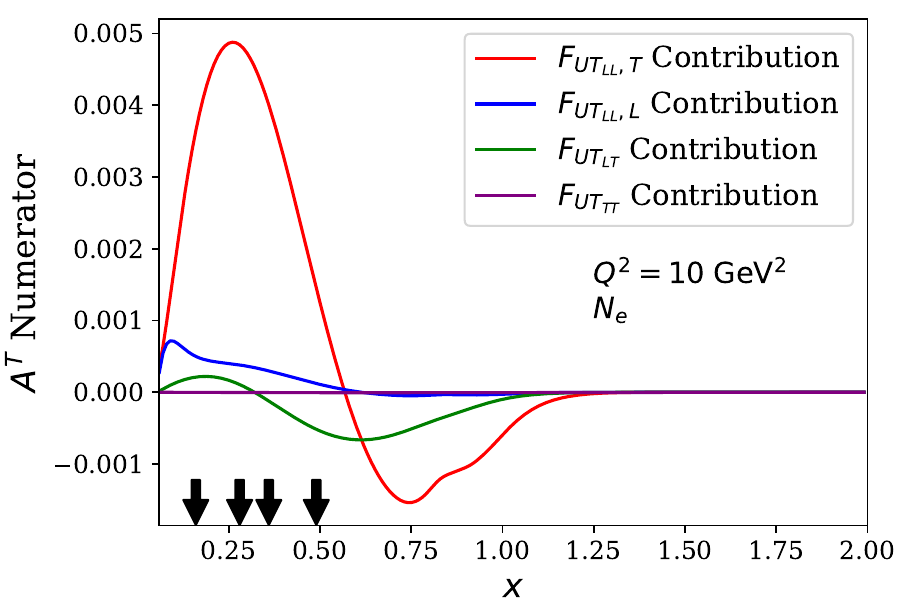}\par}    
\end{multicols}
  \caption{Contributions to the numerator of the tensor asymmetry proportional to the different tensor SF appearing in eq.\ref{eq:AT_SF}.  Curves include the polarization parameter and $\epsilon$-dependent factors.  Comparison between polarization direction along the virtual photon direction ($\bm N_q$, top row) and the electron beam direction ($\bm  N_e$, bottom row).  The convolution model calculation shown here uses the SLAC nucleon SF and the Paris deuteron wave function.}
  \label{fig:Q2Fterms}
\end{figure*}

\begin{figure*}[hbt!]
\centering
\begin{multicols}{2}
    {\includegraphics[width=0.95\linewidth,center]{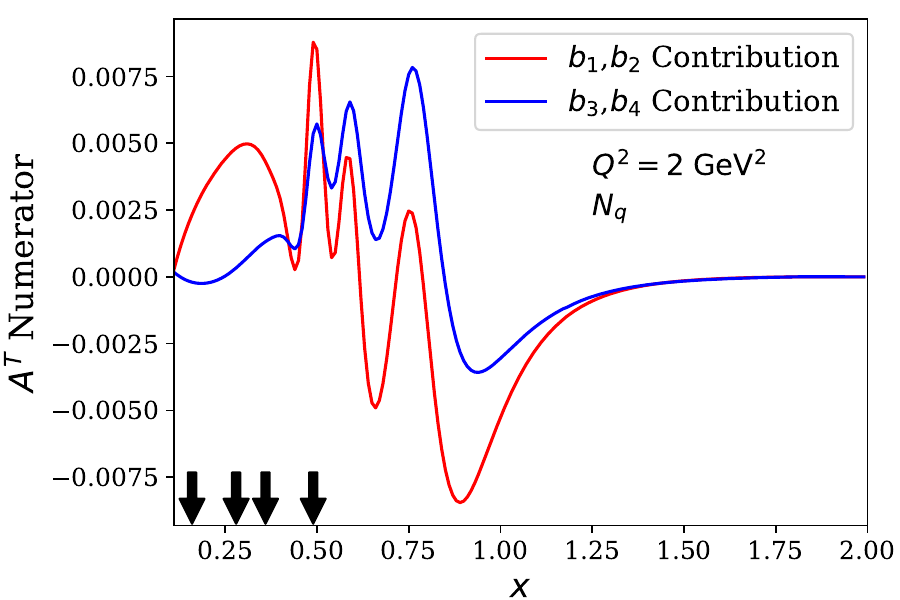}\par}
    {\includegraphics[width=0.95\linewidth,center]{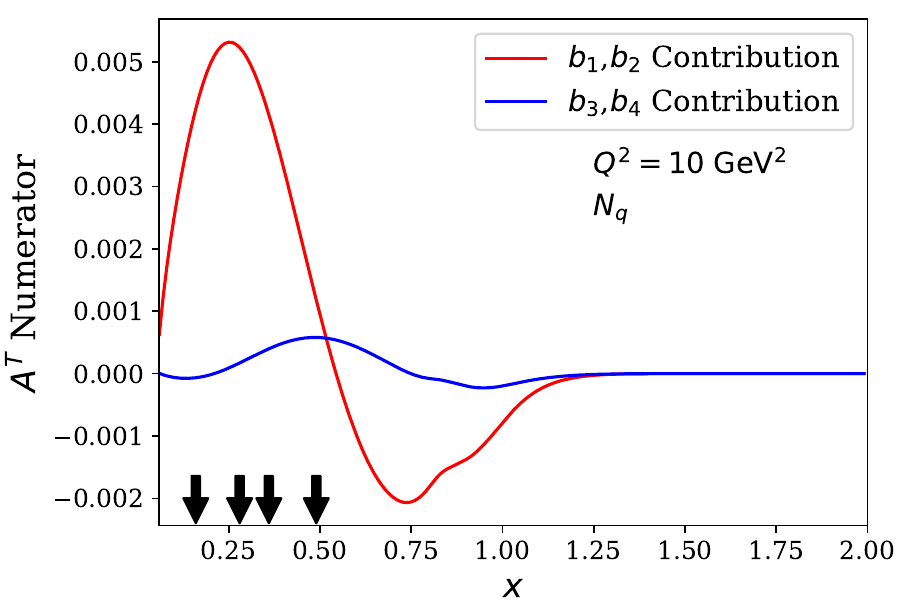}\par}
\end{multicols}
\begin{multicols}{2}
    {\includegraphics[width=0.95\linewidth,center]{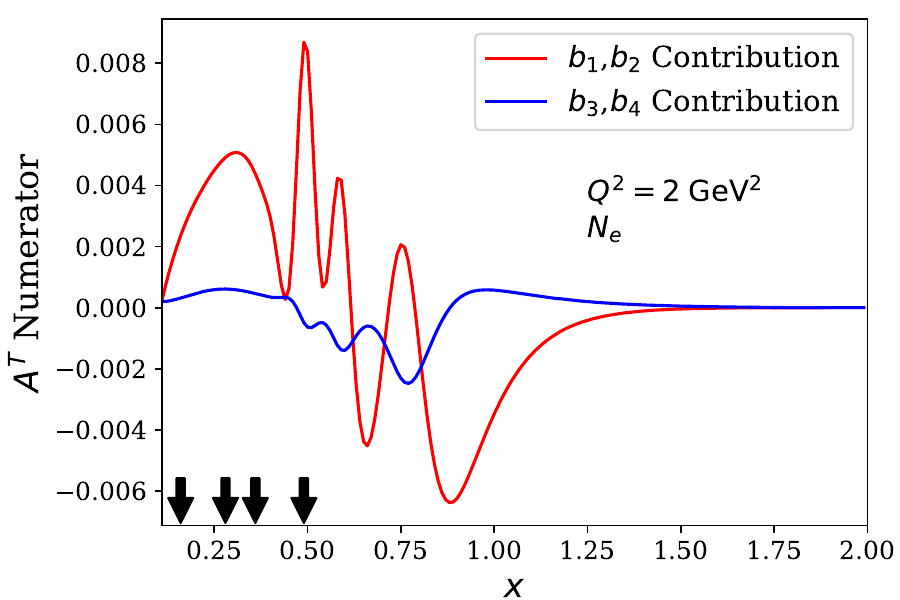}}\par
    {\includegraphics[width=0.95\linewidth,center]{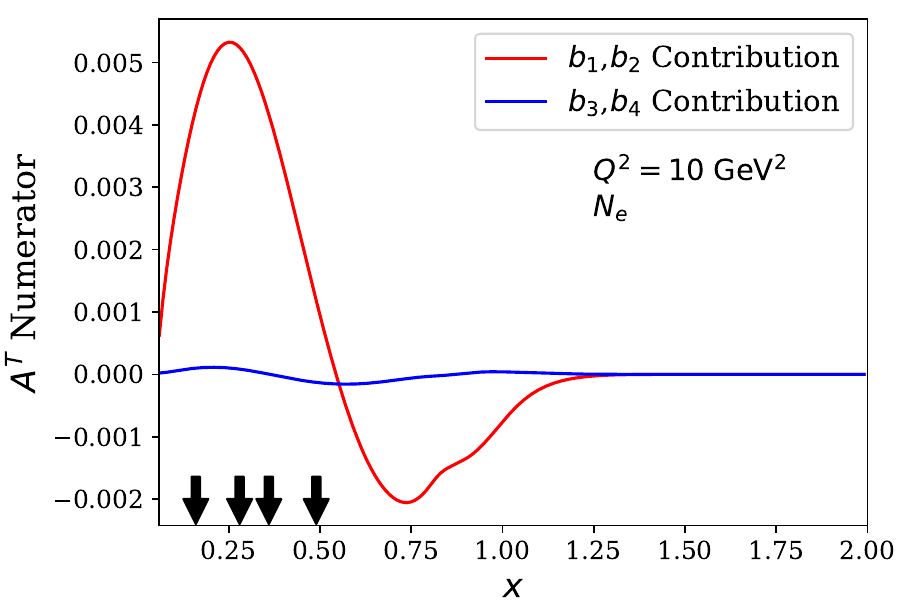}\par}    
\end{multicols}
  \caption{
    As Fig.~\ref{fig:Q2Fterms} but now the $A^T$ numerator contributions are shown with the leading twist $b_1,b_2$ terms and higher twist $b_3,b_4$ terms.
  }
  \label{fig:Q2bterms}
\end{figure*}

Using the model, we can test some of the assumptions common to all approximations of Table~\ref{table:Approxb1}.  The first is the Callan-Gross relation $b_2=xb_1$ which is valid in the Bjorken limit. Figure~\ref{fig:cgQ2} shows that, even at lower $Q^2$, using the Callan-Gross relation works quite well.  For $Q^2=2~\text{GeV}^2$ the difference is on average $\sim 20\%$ in the region of interest for the JLab experiment and becomes larger at very small $x$ or in regions where either of the structure functions has a zero crossing.  At $Q^2=10~\text{GeV}^2$ the relative difference is smaller as would be expected for this scaling limit relation and the effect of the resonances is diminished.  

Next, in Fig.~\ref{fig:Q2Fterms}, we show the contributions of the different tensor polarized SF to the numerator of the tensor asymmetry (Eq.~(\ref{eq:AT_SF})) for different polarization directions and $Q^2$ values.  We see that overall $F_{[UT_{LL},T]}$ is larger than $F_{[UT_{LL},L]}$.  The higher-twist SF $F_{[UT_{LT}]},F_{[UT_{TT}]}$ do not contribute to the asymmetry when polarization along the virtual photon direction is used, but do contribute when deuteron polarization is away from that direction, such as along the electron beam shown in the bottom panels of Fig.~\ref{fig:Q2Fterms}.  We see that $F_{[UT_{TT}]}$, which is twist 4 is negligible down to the lowest $Q^2$ value shown, while the twist-3 $F_{[UT_{LT}]}$ can be quite sizeable and can cause significant cancellations in the asymmetry when it has the opposite sign of $F_{[UT_{LL},T]}$.    

A second approximation used in Table~\ref{table:Approxb1} is neglecting the higher twist $b_3,b_4=0$ in the kinematical higher twist approximations.  The extent to which $b_3$ and $b_4$ can be neglected in the total asymmetry is shown in Fig.~\ref{fig:Q2bterms} for polarization along $\bm N_q$ and $\bm N_e$.  At $Q^2= 2~\text{GeV}^2$, we see that the contribution from $b_3,b_4$ to the asymmetry is not negligible and especially for polarization along the photon direction is on the order or larger than that of the leading twist $b_1$ and $b_2$.  For both polarization directions, as the value of $Q^2$ increases, the contributions proportional to the higher twist structure functions $b_3$ and $b_4$ decrease as expected.

\begin{figure}[hbt!]
\centering
  \includegraphics[width=\linewidth,center]{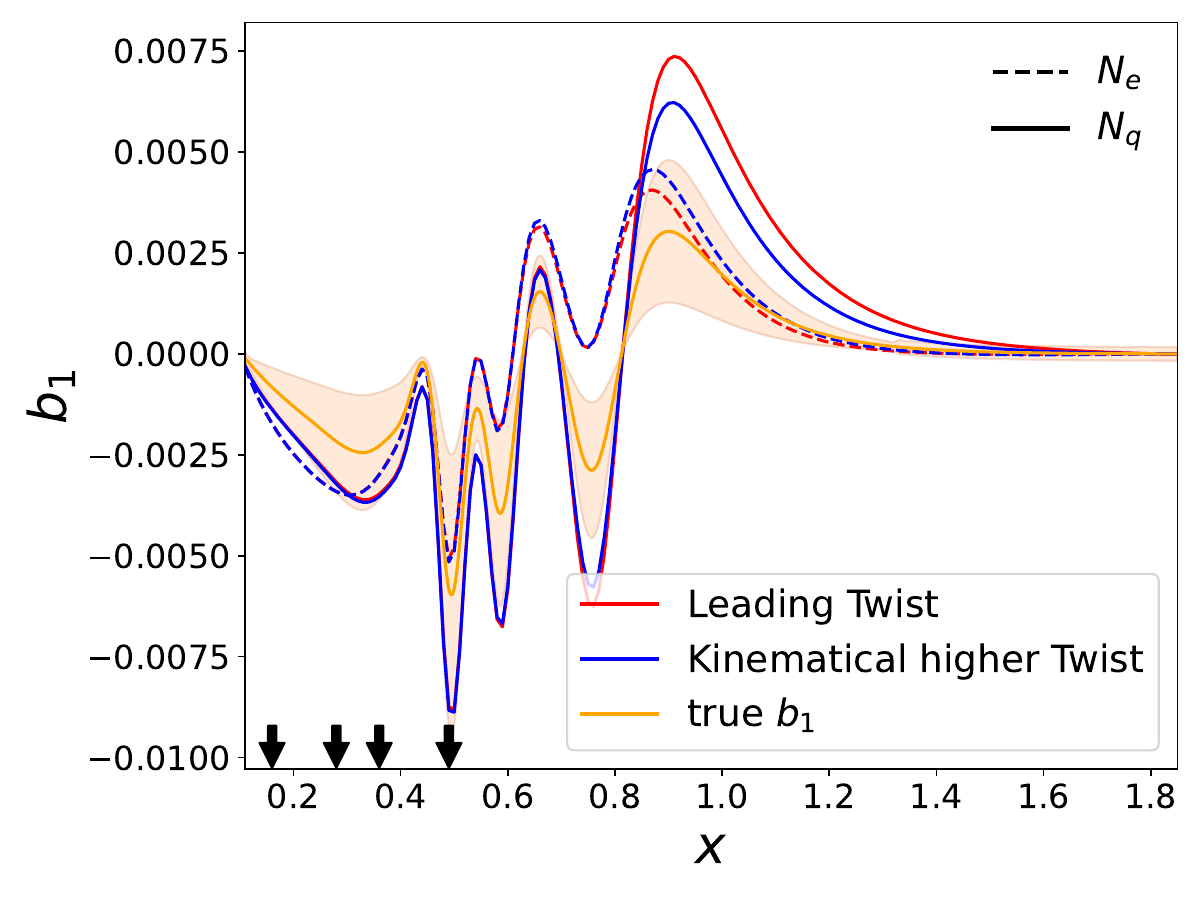}
  \caption{Comparison at $Q^2 = 2~\text{GeV}^2$ between the original model input (yellow curve) and the $b_1$ values extracted from the calculated asymmetry.  Different curves show the different polarization directions and approximations and follow the notation of Table~\ref{table:Approxb1}.  The convolution model calculation shown here uses the SLAC nucleon SF and the Paris deuteron wave function. Black arrow indicate the central $x$-values for the JLab experiment, see Fig.~\ref{fig:b1-rates}.}
  \label{fig:estimateComp}
\end{figure}

Figure \ref{fig:estimateComp} illustrates a comparison between the different extraction approximations of Table~\ref{table:Approxb1}.  Note that for the leading twist approximation we still have two different curves.  We start from asymmetries with different polarization directions ($\bm N_q,\bm N_e$) but use the same approximation relation (Eq.~(\ref{eq:AT_approx})), which results in different $b_1$ curves. While qualitatively the curves are quite similar, the oscillations due to the nucleon resonance contributions and the sign change in the structure function can result in actual sizable relative differences between an extracted $b_1$ and the ``truth'' input.  While the kinematic region of the JLab experiment avoids the region with larger resonance effects, the difference is still sizeable.  The uncertainties shown on the ``truth'' curve in Fig.~\ref{fig:estimateComp} were propagated from projected uncertainties of $A^{T}$ for the experiment~\cite{Error}. Here, we take the entire error on $b_1$ ($\sigma_{b_1}$) coming from $\sigma_{A^{T}}$, the error on the asymmetry and we can write 
\begin{equation}\label{eq:error}
    \sigma_{b_1}= \frac{\sigma_{A^{T}}}{A^{T}} b_1.
\end{equation}

\begin{figure}[hbt!]
   \centering
  \includegraphics[width=\linewidth]{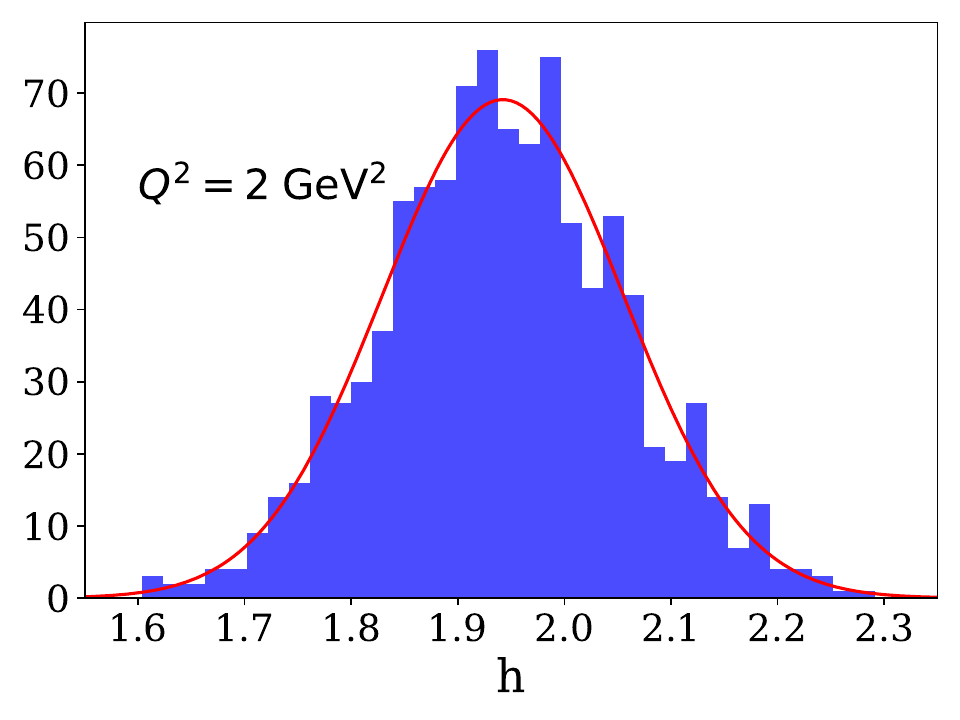}
  \caption{Distribution for $10^3$ generated h values, using the same model inputs used in Fig.~\ref{fig:estimateComp}. Here, $b_1$ was estimated using the leading twist approximation in kinematics with the polarization of the deuteron along $\bm N_q$.  The red curve shows a Gaussian fit to the histogram values.}
  \label{fig:Gauss}
\end{figure}

To quantify the systematic error between the extracted $b_1$ and model truth, we take the calculated grid from the $x$-region of interest to the experiment ($x \in [0.05,0.5]$) with interval $\Delta x=0.01$ and interpolate $10^6$ points for 1) the true $b_1$, 2) its uncertainty and 3) the extracted $b_1$ estimates that we are quantifying the error for.  This provides enough statistics for the analysis. Out of the $10^6$ interpolated points, we take random samples of $10^3$ points and calculate the following estimate
\begin{equation}\label{eq:h}
    h=\sum_i^N \frac{({b_{1i}}^{\text{true}}-{b_{1i}}^{\text{estimate}})^2}{N\sigma_{b_1,i}^2}.
\end{equation}
This corresponds to a numerical approximation of the weighted $L_2$ norm between the \emph{true} and extracted $b_1$ curve in the interval $x \in [0.05,0.5]$.
Here $N=10^3$ is the number of points in the random sample and the sum runs over the values of each variable in the sample. 
A value of $\langle h \rangle=1$ means that the systematic error induced by the approximation used in the extraction is of the same size as the other combined errors 
in the experiment. We use $10^3$ different random samples from the $10^6$ interpolated points to generate the histogram for $h$, shown in Fig.~\ref{fig:Gauss} for one choice of kinematics.  That Gaussian nature of the distribution in Fig.~\ref{fig:Gauss} is representative for all other kinematics considered in this analysis.

\begin{figure}[hbt!]
  \includegraphics[width=\linewidth]{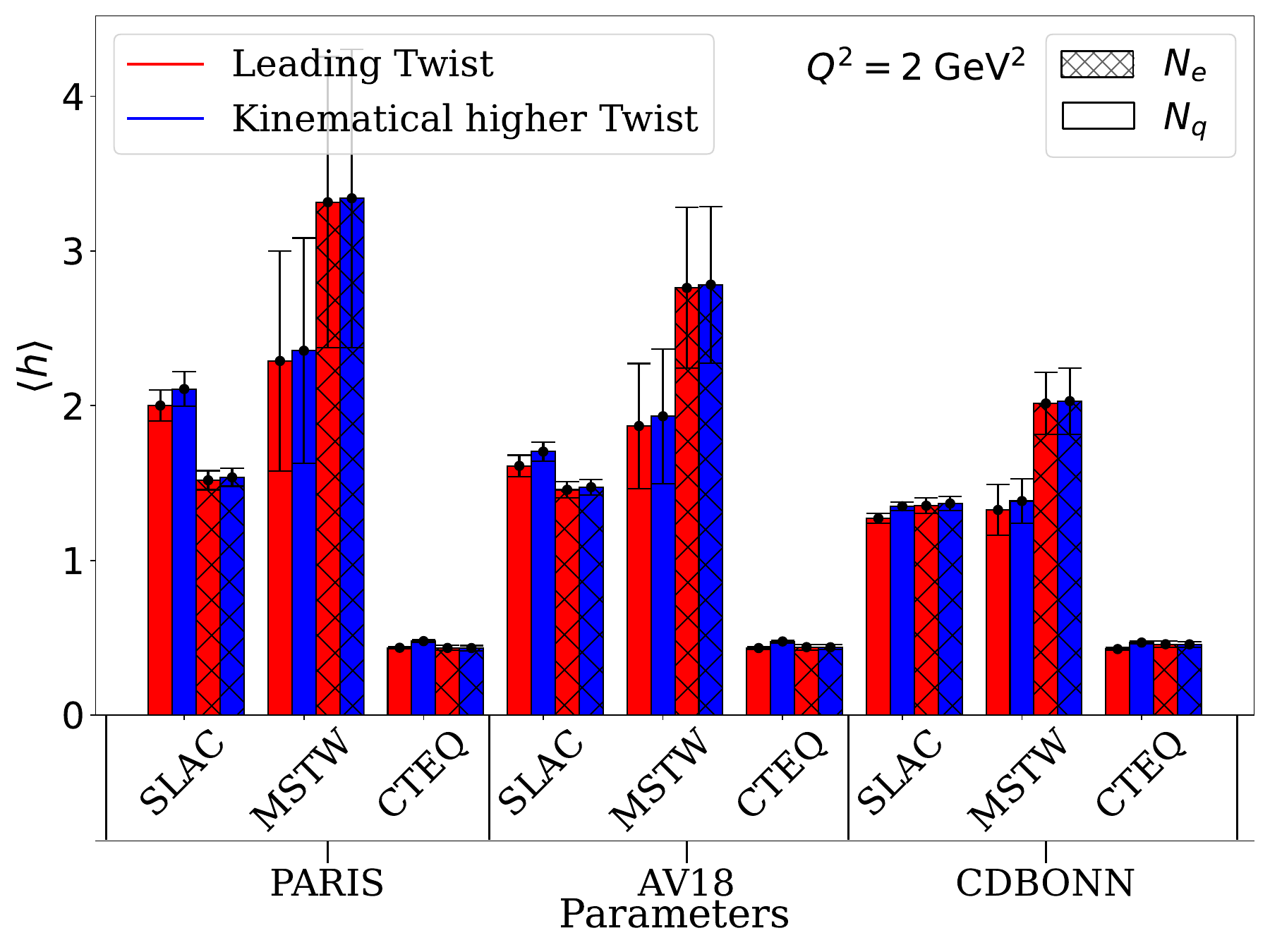}
  \includegraphics[width=\linewidth]{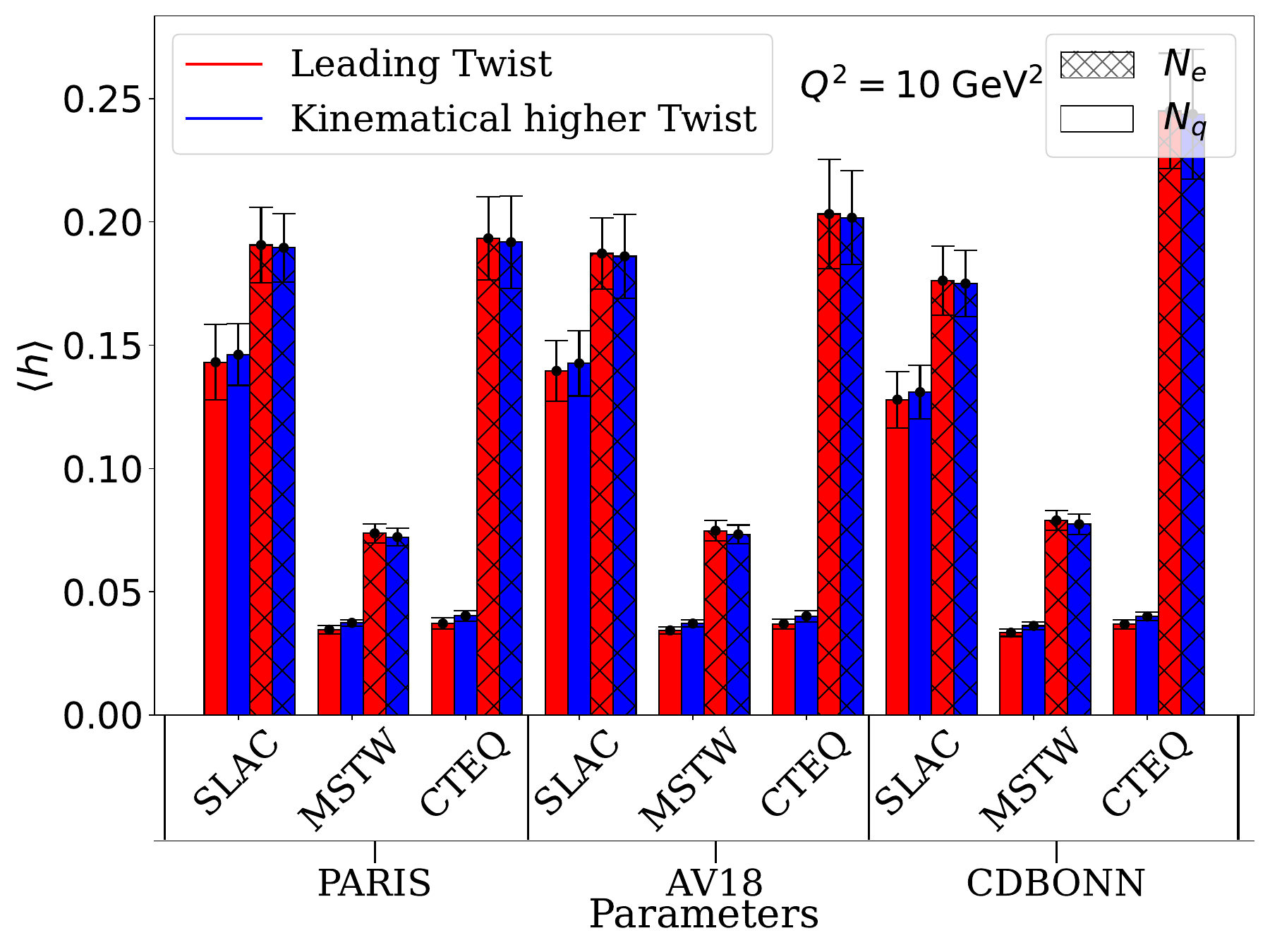}
  \caption{Average $\langle h \rangle$-values for different convolution model inputs (see horizontal axis labels) and different polarization directions and approximations (see legends and Table~\ref{table:Approxb1}). Top panel shows results for $Q^2=2~\text{GeV}^2$, the bottom panel for $Q^2=10~\text{GeV}^2$.}
  \label{fig:Qbar}
\end{figure}

To compare the error estimates, we analyze the means  of the Gaussian distributions that were fitted to the histograms, using calculations with different model inputs, polarization directions and extraction approximations. The results are summarized in Fig.~\ref{fig:Qbar}.
The interpretation of these bar charts is as follows: lower $\langle h \rangle$ corresponds to an estimate closer to the  'true' $b_1$ and a smaller systematic error. $\langle h \rangle=1$ corresponds to the error of the estimate being equal to the rest of the experimental uncertainty. As depicted in the charts, changing parameters affects which one of the approximations results in the best $b_1$ estimate.  On the one hand, we are interested if a certain polarization direction and/or approximation yields a consistently better result.  For this we can compare the different bars sharing the same parameter inputs (groups of 4).  On the other hand, we want to see trends that would be independent of the model inputs.  For that we can compare bars with different parameter labels using the same approximation (bar style and color).

By examining the two panels of Fig.~\ref{fig:Qbar}, one can observe how $\langle h \rangle$ changes with $Q^2$. At sufficiently high $Q^2$, the virtual photon direction consistently provides the better estimate.  This is in line with expectations as the $\bm N_q$ direction tensor asymmetry has only 2 tensor SF contributing, see Sec.~\ref{sec:directions}.  The average error at $Q^2=10~\text{GeV}^2$ is about a factor of 10 smaller than at $Q^2=2~\text{GeV}^2$ for the SLAC and MSTW nucleon inputs, reflecting the fact that all our considered approximations become better closer to the Bjorken limit.  At the lower $Q^2$ value, there is no consistent better polarization direction emerging.  While one would expect the photon direction to give the better result given that the asymmetry numerator contains less terms, from Fig.~\ref{fig:Q2bterms} we see that the neglected higher twist $b_3,b_4$ are much larger for polarization in the photon direction, thus worsening the used approximation again.  While there is little dependence on the deuteron wave function, the results depend quite heavily on the nucleon SF input, with the SLAC paramterization giving a smaller error for polarization along  $\bm N_e$ (for two of the three deuteron wave functions), while the CTEQ parametrization produces no significant difference and MSTW has smaller errors for the $\bm N_q$ direction.  Keep in mind that the SLAC parametrization includes resonance contributions which should produce a more realistic estimate at the JLab kinematics.  Even though a quasi-elastic nucleon $W^2>1.85~\text{GeV}^2$  cut is imposed, due to the deuteron D-wave the tensor polarized SF receive the largest contributions in the convolution integral from nucleon momenta of a few 100 MeV.  Consequently, resonance effects can still contribute in the kinematics of the experiment. The $Q^2=2~\text{GeV}^2$ results indicate that the error originating from the approximation in the extraction would be on the order of the other errors.  It is worth noting that the approximation using kinematical higher twist corrections (blue bars) never produces a significantly better error than the (simpler) leading twist approximation. 
  
\begin{figure*}[hbt!]
  \includegraphics[width=0.48\textwidth]{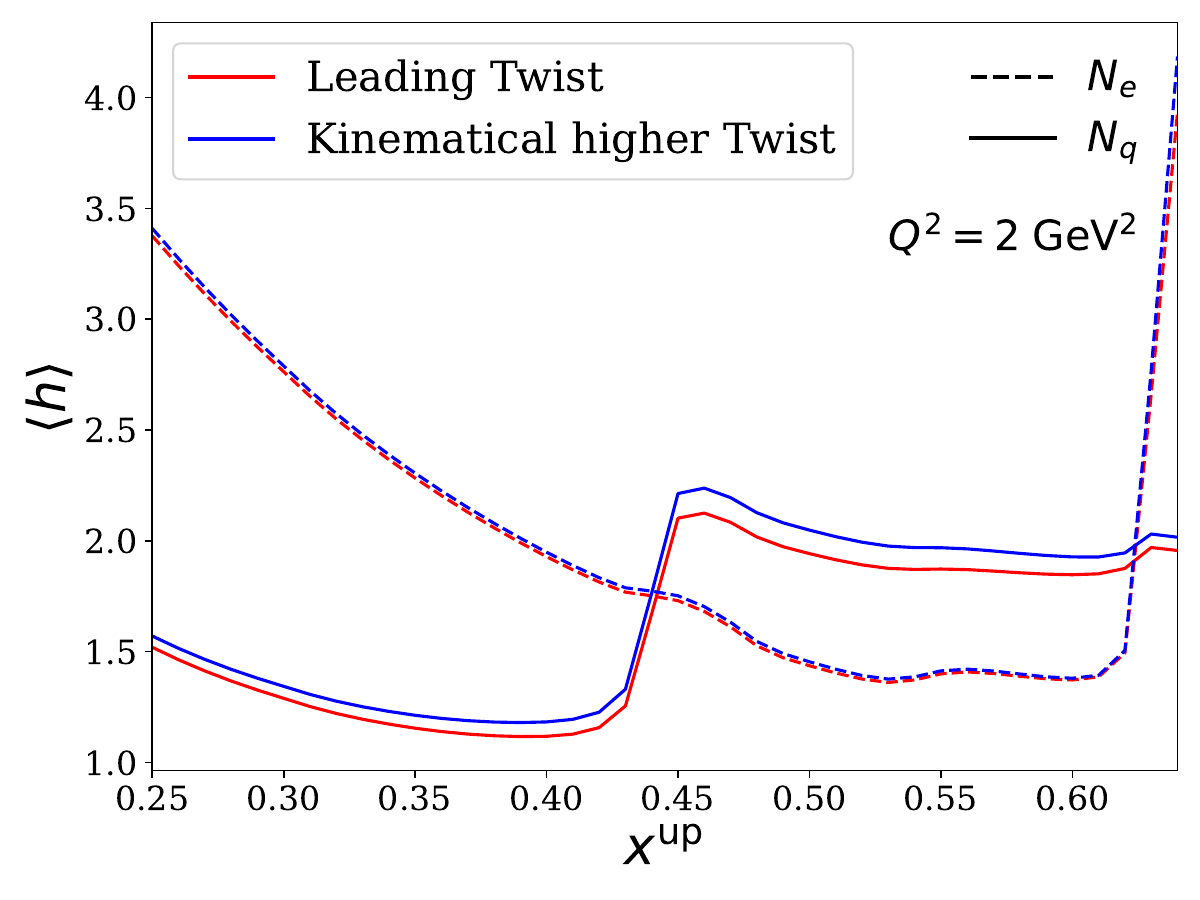}
  \includegraphics[width=0.48\textwidth]{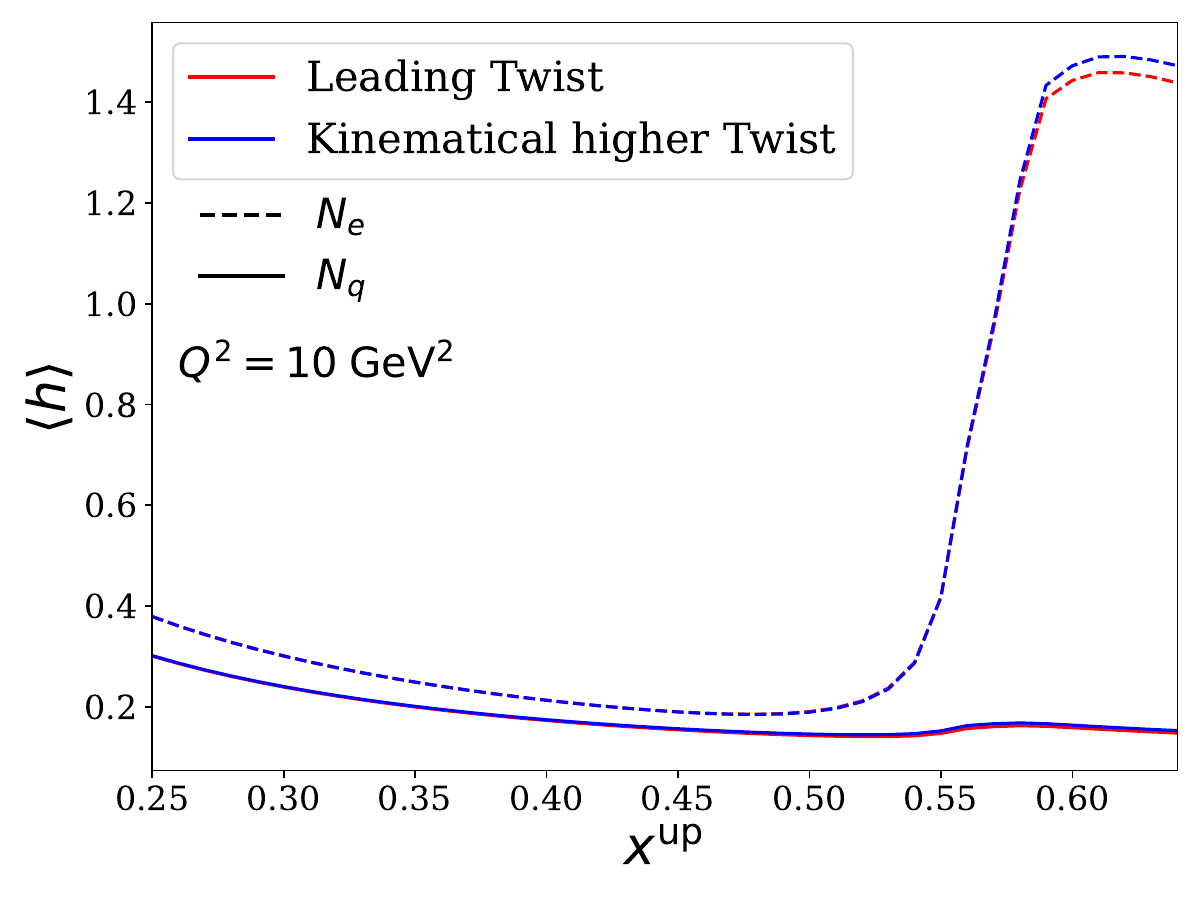}
  \caption{$\langle h \rangle$ values for different $x \in [0.05,x^{\text{up}}]$ intervals.  Left panel shows results for $Q^2=2~\text{GeV}^2$, the right panel for $Q^2=10~\text{GeV}^2$. The model inputs for plots shown here uses the SLAC nucleon SF and the Paris deuteron wave function.}
  \label{fig:upper_bound}
\end{figure*}
\begin{figure*}[hbt!]
  \includegraphics[width=0.48\linewidth]{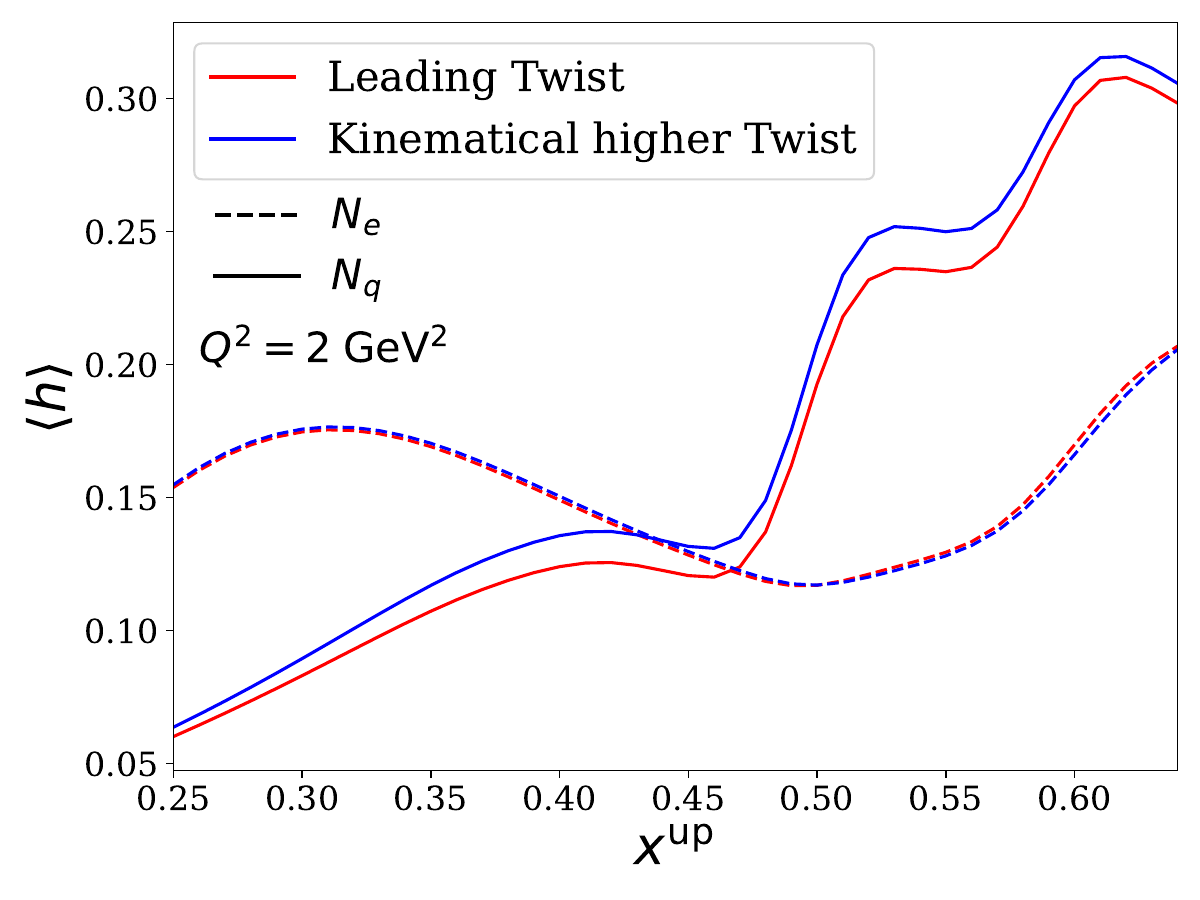}
  \includegraphics[width=0.48\linewidth]{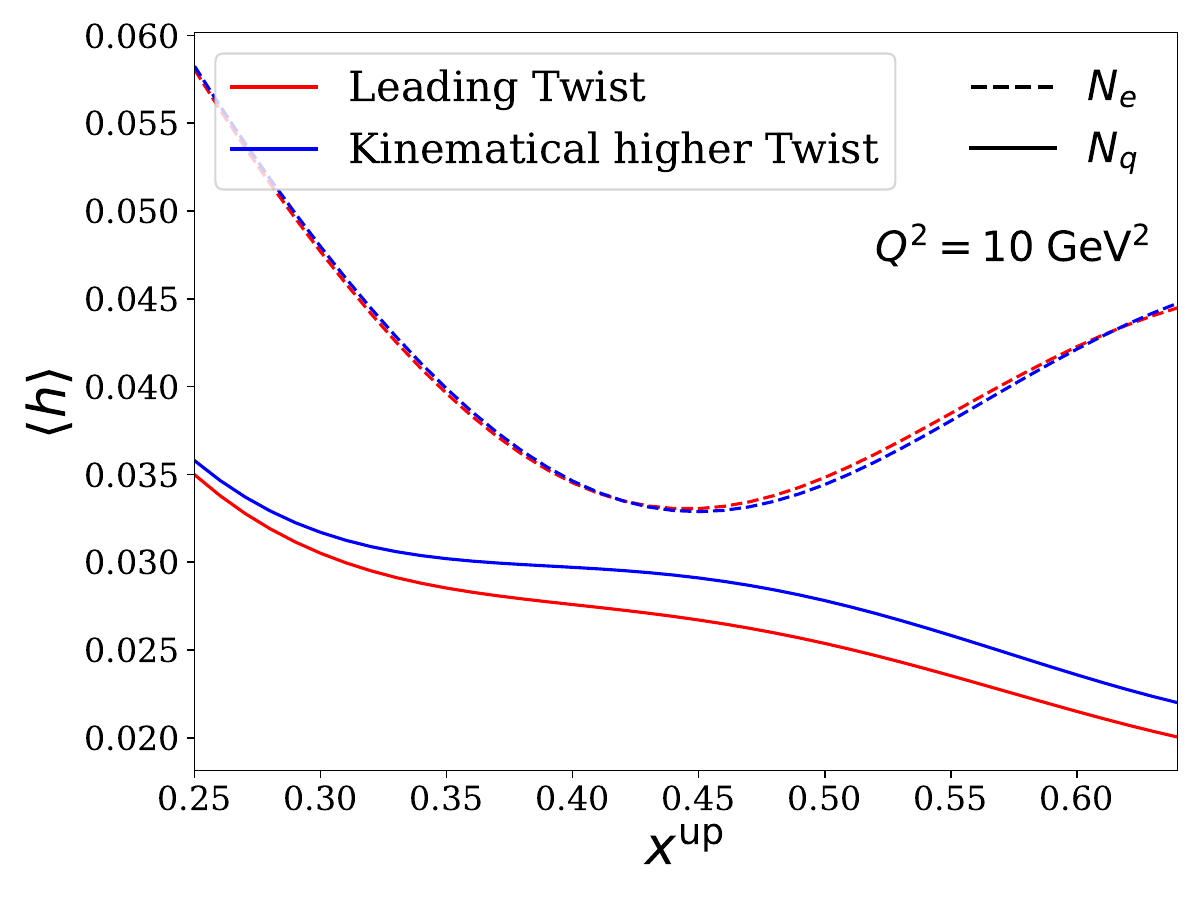}
  \caption{As in Fig. \ref{fig:upper_bound} except the uncertainty is taken to be constant throughout the entire range, see text and Eq.~(\ref{eq:ConstantError}) for details.}
  \label{fig:constant_error}
\end{figure*}

Finally, for further tests of the robustness of our results we did calculations of $\langle h \rangle$ using different $x$-intervals by varying the upper limit of the interval used in the $h$ estimation.  This is shown in Fig.~\ref{fig:upper_bound}. The approach of Eq.~(\ref{eq:error}) has the disadvantage that whenever $b_1$ crosses a zero, its error goes to zero.  This causes a large blowup in the value of $h$.  This also makes regions with smaller uncertainty on $b_1$ contribute the most to the eventual $\langle h \rangle$ values, which could bias the conclusions drawn from the results. For example, looking at Fig. \ref{fig:estimateComp} there is a zero crossing at around $x=0.43$ and this makes the $h$ values for the estimates for polarization along $\mathbf{N}_q$ blowup which can be seen in the left plot in Fig.~\ref{fig:upper_bound}.  This might lead to the conclusion that whatever curve has had less zero crossings gives the best extraction result. This happens as well for the estimates along $\mathbf{N}_e$ as can be seen around $x=0.63$ where there is another zero crossing in fig.\ref{fig:estimateComp}.  The effect is less pronounced for the high $Q^2$ calculation with $\bm N_q$ providing the better estimate across the considered intervals. A constant uncertainty weighs all events similarly in the summation of Eq.~(\ref{eq:h}) which avoids the blowup around a zero crossing.  We show results in Fig.~\ref{fig:constant_error} using a constant absolute uncertainty for which we use a value
\begin{equation}\label{eq:ConstantError}
    \sigma_{b_1}= \frac{\sigma_{A^T}}{A^T} \frac{|\min(b_1)|+|\max(b_1)|}{2}.
\end{equation}
  We see as a result that the magnitude of the error estimate is significantly smaller.
  At high $Q^2$ we see in fig.~\ref{fig:constant_error} that the virtual photon is still the better approximation and the zero crossing blowup has disappeared.   For the low $Q^2$ value, one still sees variation with the considered interval which polarization direction produces the best error.  This behavior also persists for other choices of the model inputs, not shown here.    Consequently, it is not conclusive which direction produces less error in the kinematical regions of the experiment. What is shown is that at high $Q^2$ the virtual photon direction does give less error.

\section{Conclusion}
\label{sec:concl}

An analysis was performed using a deuteron convolution model of the systematic errors induced in the extraction of $b_1$ from a single tensor polarized asymmetry using approximations.  In all scenarios, approximations that include some kinematical higher twist corrections did not perform better than the leading twist approximation.  While at larger $Q^2 \sim 10~\text{GeV}^2$ values polarizing the deuteron along the virtual photon direction provides a smaller error, at the kinematics relevant to the JLab experiment at lower values of $Q^2$, there is no clear preference between polarizing along the electron beam or virtual photon direction.  Different model inputs yield different results, with the magnitude of the error depending on how the error on $b_1$ was treated.  This demonstrates that i) this is an effect that should be included in a proper error analysis, ii) there is no clear preference at low $Q^2$ in which direction to polarize the deuteron to minimize this error, with the electron beam the more practical choice in the fixed target JLab setup.  
While these conclusions are based on calculations within one model, it would be interesting to do a similar exercise using a global analysis framework to quantify an error estimate using closure tests.  Using measurements with different tensor polarizations at the same kinematics could reduce the uncertainties and in principle allow extraction of all independent structures (See Ref. \cite{JLABt20:2000uor} for a similar approach using polarization transfer in elastic $ed$ scattering).

\section*{Acknowledgements}
We want to thank Elena Long for useful discussions. WC is supported through NSF grants PHY-2111442 and PHY-2239274, AS acknowledges support through a Bridge to Doctorate NSF fellowship at FIU.  This work has been supported by EURO-LABS ``EUROpean Laboratories for Accelerator Based Science'' which received funding from the Horizon Europe research and innovation programme under grant agreement No. 101057511. This material is also based upon work supported by the U.S. Department of Energy, Office of Science, Office of Nuclear Physics under grant DE-FG02-88ER40410. We used Jaxodraw \cite{Binosi:2008ig} for Fig. \ref{fig:DIS}.

%
\bibliographystyle{spphys}
\bibliography{bibliography.bib}

\end{document}